\documentclass[apl,aip,onecolumn,preprint,superscriptaddress]{revtex4-2}

\usepackage{bm}
\usepackage[colorlinks=true,linkcolor=blue,citecolor=blue]{hyperref}
\usepackage{times}
\usepackage{amsmath}
\usepackage{amssymb}
\usepackage{amsthm}
\usepackage{amsfonts}
\usepackage{enumerate}
\usepackage{latexsym}
\usepackage{ifpdf}
\newcommand{\tsub}{\textsubscript}
\newcommand{\beq}{\begin{equation}}
\newcommand{\eeq}{\end{equation}}
\usepackage{graphicx}
\usepackage{makeidx}
\usepackage{color}
\usepackage{mhchem}
\usepackage{gensymb}

\begin{document}

\title{Electronic and magnetic properties of  epitaxial thin film of Nd\tsub{0.5}Ba\tsub{0.5}MnO\tsub{3} }
 \author {Siddharth Kumar}
\affiliation  {Department of Physics, Indian Institute of Science, Bengaluru  560012, India}
\author {Shashank Kumar Ojha}
\affiliation  {Department of Physics, Indian Institute of Science, Bengaluru 560012, India}
\author {Ranjan Kumar Patel}
\affiliation  {Department of Physics, Indian Institute of Science, Bengaluru 560012, India}
\author {Prithwijit Mandal}
\affiliation  {Department of Physics, Indian Institute of Science, Bengaluru 560012, India}
\author {Nandana Bhattacharya}
\affiliation  {Department of Physics, Indian Institute of Science, Bengaluru 560012, India}
\author {S. Middey}
\email{smiddey@iisc.ac.in}
\affiliation  {Department of Physics, Indian Institute of Science, Bengaluru 560012, India}

\begin{abstract}
Contradictory reports about the electronic and magnetic behavior of bulk Nd\tsub{0.5}Ba\tsub{0.5}MnO\tsub{3} with uniform disorder exist in literature. In this work, we investigated single crystalline, A-site disordered thin films of Nd\tsub{0.5}Ba\tsub{0.5}MnO\tsub{3}, grown  on SrTiO\tsub{3} substrate by pulsed laser deposition. The  epitaxial growth of these films in the layer-by-layer fashion has been confirmed by \textit{in-situ} reflection high energy electron diffraction, atomic force microscopy, X-ray reflectivity, and X-ray diffraction measurements. Interestingly, these films are found to be electrically insulating and exhibit spin-glass behavior at low temperatures. This offer a distinct approach to study Ba based half-doped manganites and opportunity to further   manipulate competing charge/orbital ordering and ferromagnetism in such systems, through heterostructure engineering.
\end{abstract}

\maketitle

Hole doped manganites ($RE_{1-x}AE_x$MnO$_3$ where $RE$ and $AE$  correspond to a rare-earth and alkaline-earth metal, respectively) exhibit several interesting phenomena like colossal magnetoresistance (CMR), metal-insulator  transition (MIT), charge/orbital ordering (CO/OO), several types of spin ordering, etc., as a function of carrier doping and electronic bandwidth~\cite{Imada:1998p1039,Dagotto:2001p1,Tokura:2006p797}. The behavior of these phases are highly tunable under the application of  external perturbations such as magnetic field, electric field, pressure, strain, light, etc.~\cite{Dagotto:2001p1,Tokura:2006p797,Imada:1998p1039,Kuwahara:1995p961,Hwang:1995p914,Miyano:1997p4257,Asamitsu:1997p6637,Ward:2009p885,Beaud:2014p923,Zhang:2016p956}. The parent member ($RE$MnO$_3$) of the series  contains Jahn-Teller active Mn$^{3+}$ ($d^4$: $t_{2g}^3, e_g^1$) ions and the Mn$^{3+}$-O- Mn$^{3+}$ superexchange interaction is antiferromagnetic according to Goodenough-Kanamori rule\cite{Goodenough:prb1955p564}. The partial replacement of $RE^{3+}$ by $x$ amount of $AE^{2+}$ results in $x$ amount of Jahn-Teller inactive Mn$^{4+}$ ions, together with  ferromagnetic Mn$^{3+}$-O-Mn$^{4+}$ double exchange interaction. Thus, the chemical doping introduces randomness in the potential energy, Mn valency, orbital configuration, magnetic interaction, etc. Hence, the disorder is expected to play  significant role in the physics of CMR manganite. This issue has been examined by synthesizing ordered and disordered structure [see Fig. 1(a)] of half doped manganite $RE_{0.5}$Ba$_{0.5}$MnO$_3$~\cite{Akahoshi:2003p177203,Yamada:2017p035101,Trukhanov:2002p184424,Nakajima:2004p2283}.

  \begin{figure}
	\includegraphics[scale=0.45]{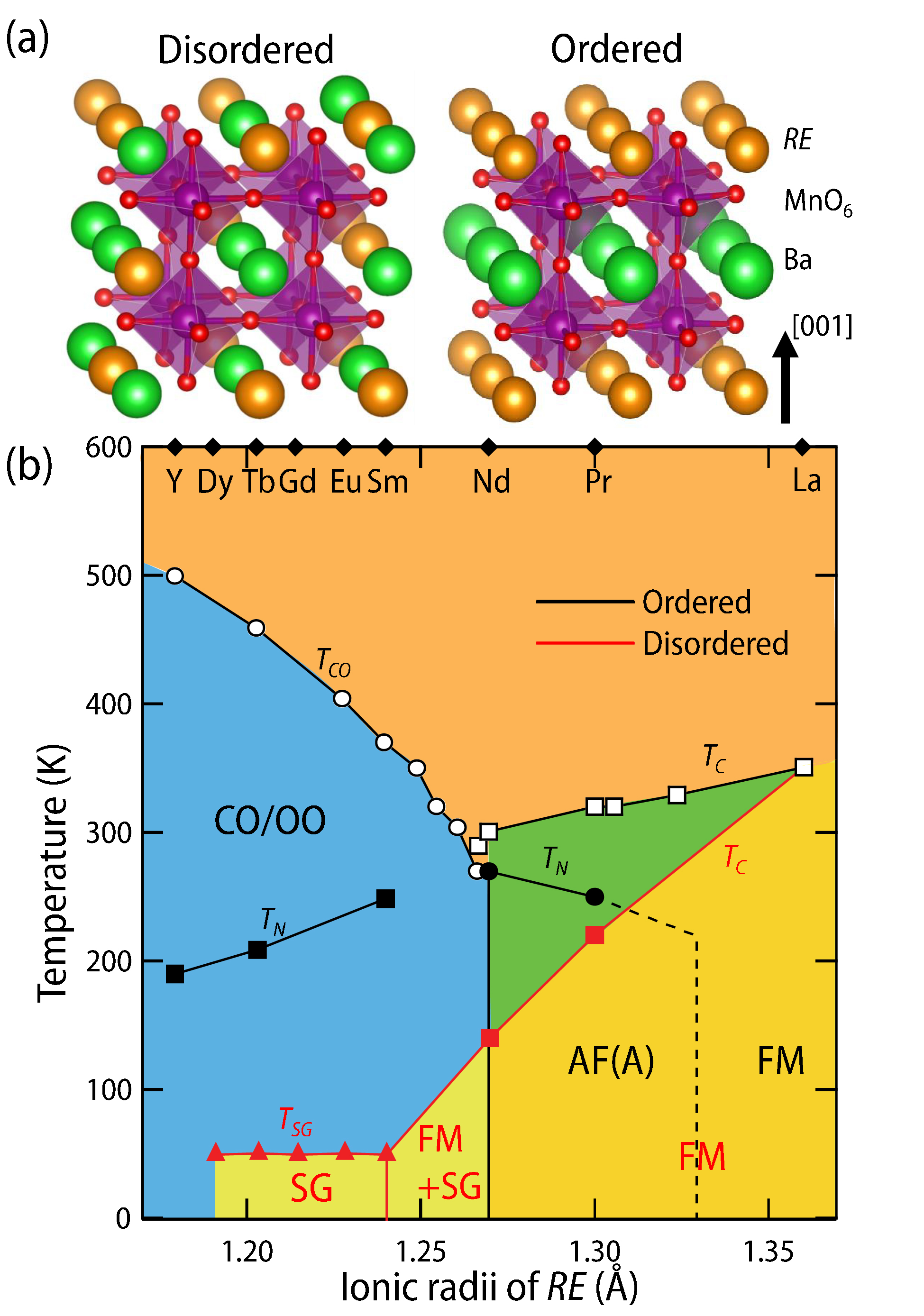}
	\caption{\label{fig:PhaseDia}(a) Disordered and ordered structure of $RE_{0.5}$Ba$_{0.5}$MnO$_3$. (b) Phase diagram of $RE_{0.5}$Ba$_{0.5}$MnO$_3$ as a function of $RE$.  Reproduced with permission from Phys. Rev. Lett. {\bf 90}, 177203 (2003). Copyright 2003 American Physical Society. SG and AF(A) represent spin-glass states, and $A$-type antiferromagnetic phase, respectively.}
\end{figure}

The phase diagram of ordered and disordered samples of $RE_{0.5}$Ba$_{0.5}$MnO$_3$ are drastically different, as shown in Fig. \ref{fig:PhaseDia} and summarized in Ref.~\onlinecite{Akahoshi:2003p177203}. Within this series, Nd$_{0.5}$Ba$_{0.5}$MnO$_3$ (NBMO) is a very special case as it is located extremely near  the  boundary between the  competing CO/OO and ferromagnetic (FM) phases. Interestingly,  observed properties of NBMO are strongly dependent on the preparation methods.
 Polycrystalline NBMO sample with Nd/Ba ordering along the $c$ direction undergoes an A-type antiferromagnetic transition around 270 K with a steep change in resistance. Both phases across the transition are insulating ($d\rho/dT< $0)~\cite{Akahoshi:2003p177203}. Whereas, the single crystalline, ordered NBMO sample undergoes a metal to insulator transition around 290 K and an antiferromagnetic transition at 235 K~\cite{Yamada:2017p035101,Yamada:2019p126602}. While the single crystal of disordered NBMO shows an MIT and ferromagnetic transition at 150 K~\cite{Akahoshi:2003p177203}, the disordered, polycrystalline sample remains insulating below room temperature and undergoes a spin glass transition around 50 K~\cite{Trukhanov:2002p184424,Nakajima:2004p2283}.

 Epitaxial stabilization is another route to obtain single crystalline sample, which further offers several ways (e.g. strain engineering, interface engineering, geometrical lattice engineering, etc.) to tune the ground state of complex oxides~\cite{Schlom:2008p2429,Hwang:2012p103,Chakhalian:2014p1189,Middey:2016p305}. Surprisingly, thin film work on Ba-based half-doped manganite is very limited~\cite{Nakajima:2007p5355} and the electronic and magnetic behavior of single crystalline NBMO film has not been reported so far. A systematic investigation about the fabrication of NBMO film will also provide crucial guidelines for the epitaxial growth of other members of  the $RE_{0.5}$Ba$_{0.5}$MnO$_3$ series.

 In this work, we report on the layer by layer growth of disordered NBMO film on  SrTiO$_3$ (STO) (0 0 1) substrate by the pulsed laser deposition technique (PLD). The epitaxial strain is expected to be small ($\epsilon \sim$ +0.23\%) as the lattice constant of disordered NBMO (3.896\AA ~\cite{Trukhanov:2002p184424}) is close to that of  STO (3.905 \AA).    A combined characterization using \textit{in-situ} reflection high energy electron diffraction (RHEED), atomic force microscopy (AFM), X-ray diffraction (XRD), X-ray reflectivity (XRR) established the epitaxial growth of the film with very smooth surface morphology, excellent crystallinity and sharp film/substrate interface structure. We have found that the epitaxial NBMO films are insulating below room temperature and the application of an external magnetic field (up to 9T) does not result in any transition to a metallic phase.
  DC magnetic measurements have found that the film undergoes a spin-glass transition around 24 K.

 Epitaxial thin films of NBMO were grown on single terminated STO (001) substrates by   a PLD system. To prepare a single terminated surface, as received STO single crystals (5$ \times $5$\times$0.5 mm$^{3}$) from Shinkosa, Japan were cleaned ultrasonically, with acetone and methanol and heated at 950$^{\circ}$ C for 1 hour within a box furnace. Afterward, the substrates were washed at room temperature with DI water, acetone and methanol, consecutively using an ultrasonicator. For the film deposition, a KrF excimer laser ($\lambda$ = 248 nm) at a repetition rate of 2 Hz was used to ablate a ceramic target of disordered NBMO. NBMO films, 25 uc in thickness, were grown at 870$^\circ$ C under a dynamic oxygen pressure of 150 mTorr with \textit{in-situ} RHEED monitoring.  After that, the  films were capped with few layers of STO at the same deposition condition to avoid any possible over oxidation of the film once exposed to the atmosphere.   We  also grew a 100 uc thick NBMO film and results of that sample have been shown in  supplementary material. The surface morphology of the bare substrates and films were investigated using a Park systems AFM. A lab-based Rigaku Smartlab diffractometer was used for XRD and XRR measurements. DC magnetic measurements were carried out using a Quantum Design SQUID magnetometer. Electrical resistance measurements were carried out by standard four-probe van der Pauw technique in an Oxford Integra LLD system using a Keithley 2450 source measure unit. Ohmic contacts were made using ultrasonic bonding of aluminum wire. The upper limit of resistivity measurement in our setup is  $\sim$ 5 $\Omega$-cm.

\begin{figure}
	\includegraphics[scale=0.35]{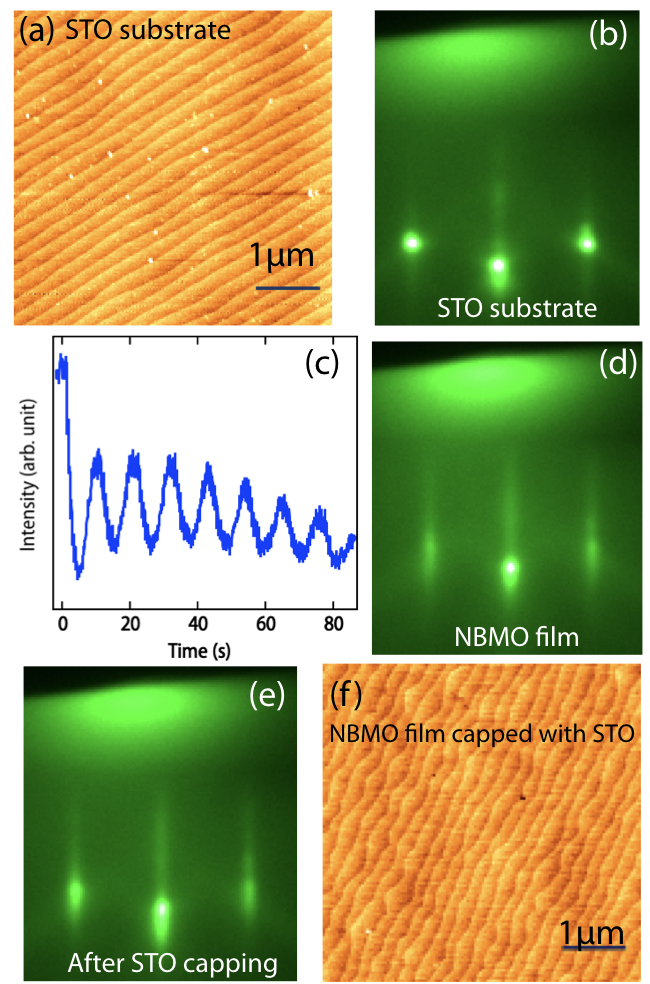}
	\caption{\label{fig:STR} (a) AFM surface morphology and (b) RHEED image of single terminated STO (0 0 1) substrate. The   oscillation in the intensity of the RHEED specular  spot during the deposition of NBMO layers signifies layer by layer growth. RHEED pattern after completion of the deposition of (d) NBMO and (e) STO capping layers. (f) AFM surface morphology of STO capped NBMO film.	}
\end{figure}

First, we discuss the morphological and structural characterization of the treated substrate and the grown film. Fig. \ref{fig:STR}(a) shows an AFM image of the treated substrate, which clearly shows the presence of large atomically flat terraces with width about 200 nm and one unit cell height. The surface roughness within a terrace is around 150 pm. The sharp specular (0 0) and off-specular (0 $\pm$1) RHEED spots  (Fig. \ref{fig:STR}(b)) also confirms the smooth surface morphology of the substrate. The intensity of the (0 0) reflection was recorded to monitor the progress of the growth. The desired layer-by-layer growth of the NBMO film is confirmed by the observation of a prominent oscillation of the RHEED intensity, plotted in Fig. \ref{fig:STR}(c). Streaky RHEED patterns (Fig. \ref{fig:STR}(d)-(e)) establish a very smooth surface morphology of the film (both before and after the deposition of STO capping layers). The retention of atomically flat terraces with unit cell height in AFM imaging of the NBMO/STO film (Fig. \ref{fig:STR}(f)), similar to the STO substrate, further demonstrates the excellent surface morphology of the sample.

\begin{figure*}
    \centering
    \includegraphics[scale=0.7]{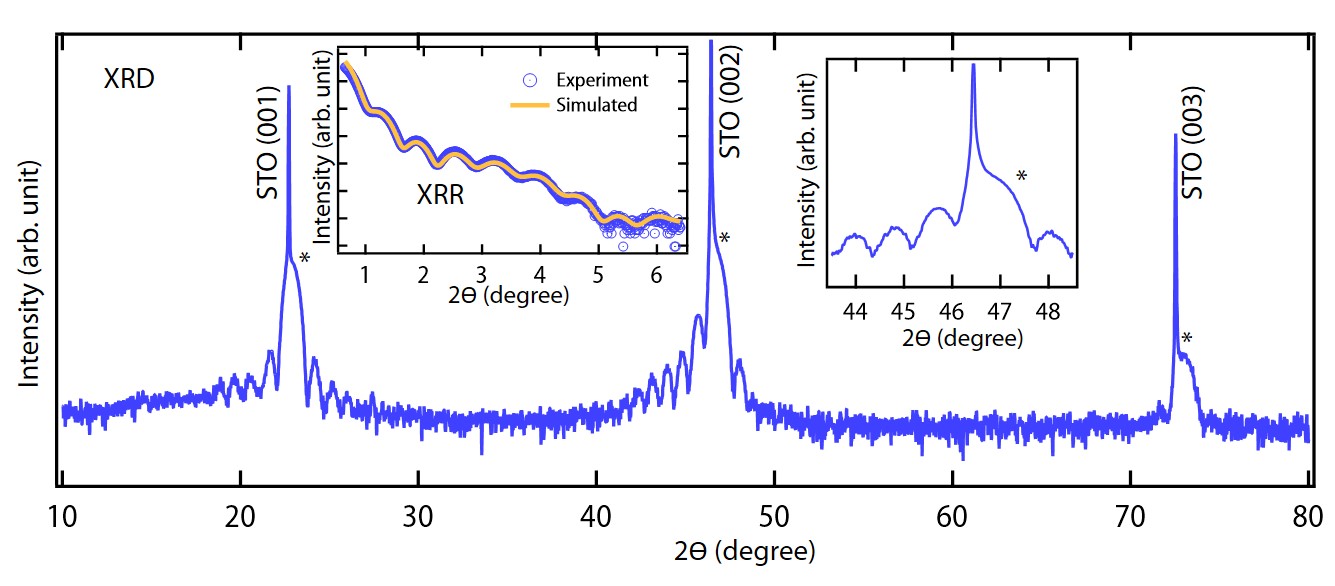}
    \caption{Long range $\omega$-$2\theta$ scan along [00l] direction of NBMO film; (left inset) shows the XRR of the same film along with the fit; (right inset) zoomed view of [002] peak, here the film peak (*) and the Kiessig fringes can be seen clearly.}
    \label{fig:XRD}
\end{figure*}

More information about the film thickness and interface roughness have been obtained by XRR. Left inset of Fig. \ref{fig:XRD} shows an experimentally observed reflectivity curve along with a simulated pattern using GenX program ~\cite{Bjorck:2007p1174} for a 25 uc NBMO film, capped with  STO. The observation of the intensity oscillations over the whole scan range indicates a highly crystalline nature and smooth interfaces. Our XRR fitting finds the NBMO layer thickness is around 98 \AA\  and the interfacial roughness is less than 3 \AA.  The epitaxial nature of the NBMO layers on STO can be seen from $\omega$-$2\theta$ XRD scans [Fig.~\ref{fig:XRD}], which consists of sharp substrate peaks, broad film peaks (denoted by $\star$) (right inset of Fig. \ref{fig:XRD}), and a set of Kiessig fringes. The out of plane lattice constant ($c$ = 3.885 \AA) is smaller than the bulk lattice constant, as expected for a tetragonal distortion under tensile strain.  The  film thickness  obtained from XRD  (25$\times c$) for this NBMO film is very close to the value found from XRR analysis.

\begin{figure}
    \centering
    \includegraphics[scale=.4]{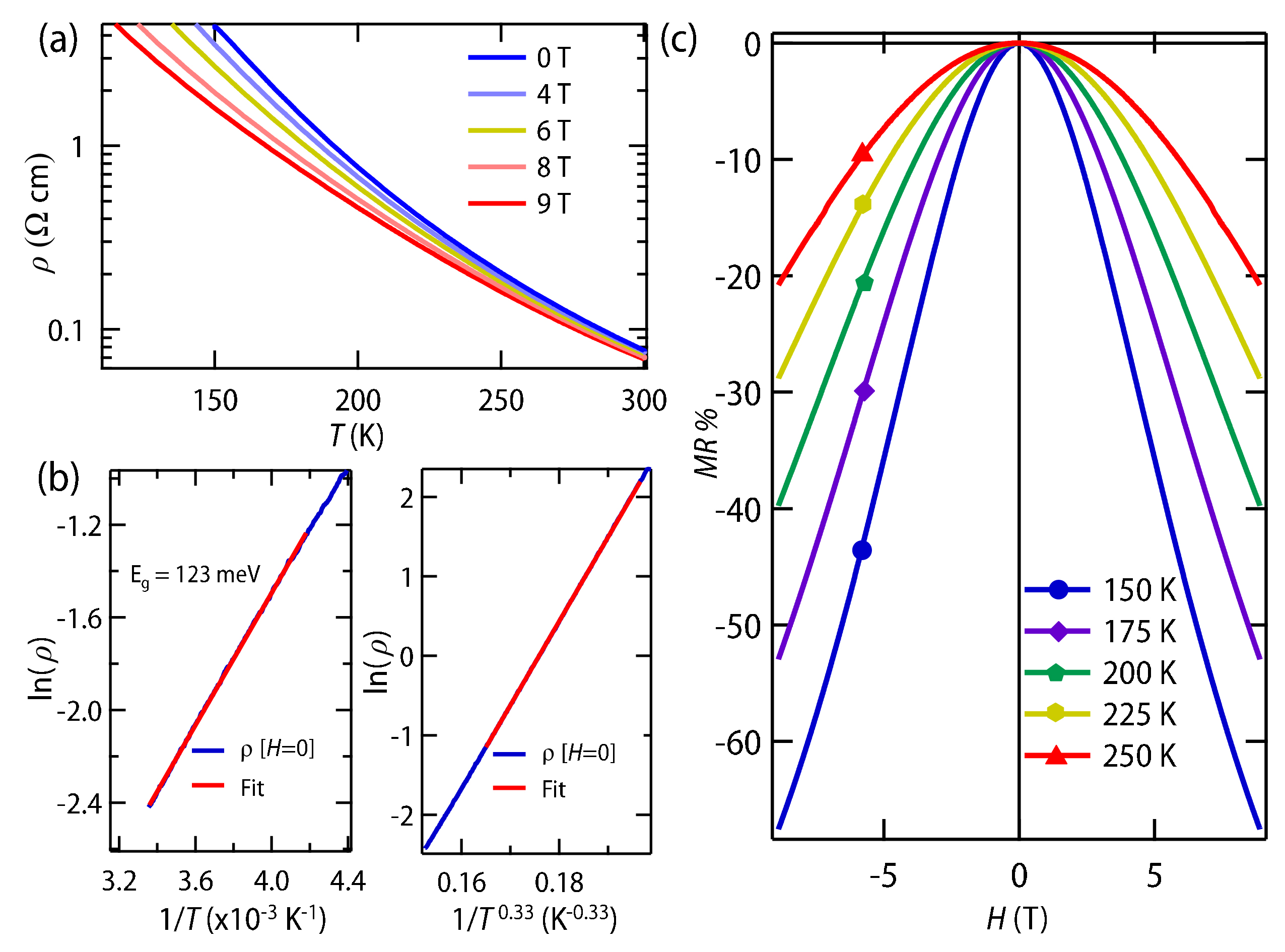}
    \caption{(a) Resistivity vs temperature plots for various applied magnetic fields (b) left and right panels show the activated behaviour and 2D VRH nature of resistivity respectively (c) MR at various temperatures up to $\pm$9 T field. }
    \label{fig:RT-MR}
\end{figure}

Having confirmed the excellent structural quality of the films, we investigated  the electrical and magnetic properties of these films. As shown in the Fig.~\ref{fig:RT-MR}(a), the 25 uc film is insulating within the temperature range of our measurement (150 K-300 K).  Under zero-magnetic field, $\rho$  shows an activated behaviour within $T$= 300 K to 240 K as shown in left panel of Fig. \ref{fig:RT-MR}(b). We have found the    activated  gap to be $\sim$123 meV, which is very close  to the value (131 meV) obtained for a polycrystalline, disordered NBMO sample  ~\cite{Nakajima:2004p2283}. Below 240 K, the resistivity follows
 two-dimensional Mott variable range hopping (VRH)  (right panel of Fig.~\ref{fig:RT-MR}(b)).  The 100 uc film also exhibits similar electrical transport (shown in supplementary materials).

The application of a magnetic field ($H$) results in a lower $\rho$ but does not induce a metallic phase (Fig.~\ref{fig:RT-MR}(a)). We have also measured $\rho$ as a function of $H$ at several fixed temperatures.  The variation of magnetoresistance (MR)  [defined as
 $\mathrm{MR}(H) \% = \frac{\rho[H]-\rho[H=0]}{\rho[H=0]}\times100\%$] as a function of $H$ have been plotted in Fig.~\ref{fig:RT-MR}(c). The negative MR is a very common feature of $RE_{1-x}AE_x$MnO$_3$~\cite{Tokura:2006p797}, and ranges from few to multiple orders of magnitude, depending on whether it is related to double-exchange mechanism (increase of electronic bandwidth due to alignment of Mn spins in an external magnetic field)~\cite{Urushibara:1995p14103} or field induced melting of charge-ordering (CO)~\cite{Tomioka:1996pR1689}. The second scenario is discarded for disordered $RE_{0.5}$Ba$_{0.5}$MnO$_3$ compounds with lower bandwidth as the uniform disorder breaks the CO/OO correlation down to nanometer scale~\cite{Mathieu:2004p227202}.

\begin{figure}
    \centering
    \includegraphics[scale=.65]{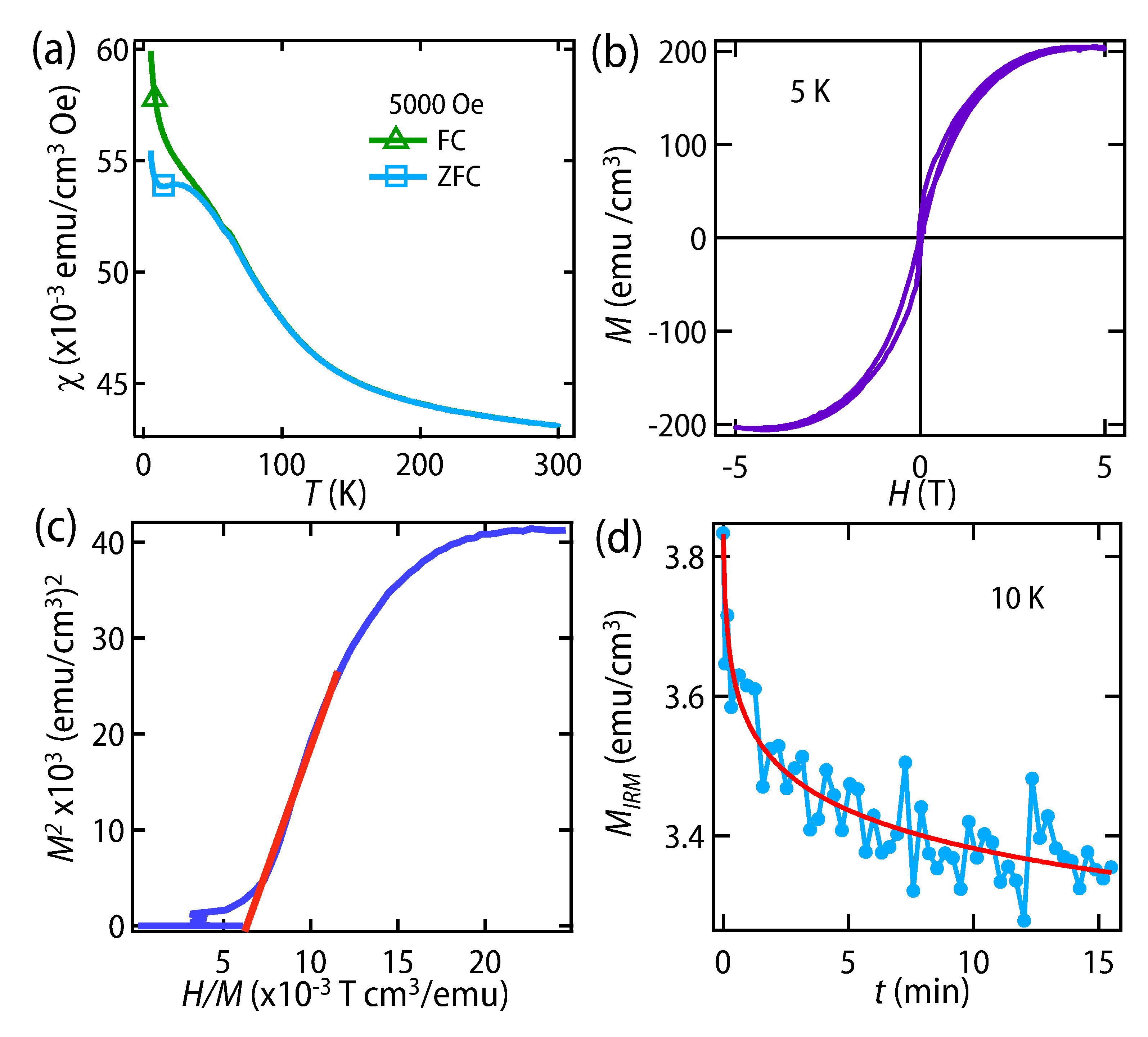}
    \caption{(a) Magnetic susceptibility vs temperature (b) $M-H$ curve at 5 K (c) Arrrot plot corresponding to the $M-H$, the red line is a guide to the intercept of the curve to the x-axis  (d) isothermal remnant magnetisation at 10 K along with the logarithmic relaxation fit as discussed in the text.}
    \label{fig:MT-MH}
\end{figure}

Disordered $RE_{0.5}$Ba$_{0.5}$MnO$_3$ with short-range CO/OO exhibits spin-glass behavior~\cite{Mathieu:2004p227202}.  Fig~\ref{fig:MT-MH} (a) shows the temperature dependence of zero-field cooled (ZFC) and field cooled (FC) susceptibility ($\chi$). For the ZFC (FC) measurement, the sample was cooled from 300 K to 5 K in the absence of an external magnetic field (in presence of $H$ = 5000 Oe)  and the magnetization was recorded during the heating run under the application of $H$ = 5000 Oe in both cases. A constant diamagnetic background arising from the STO substrate is subtracted for evaluating $\chi$ of NBMO film. We observe a bifurcation of ZFC-FC curves near 45 K and a cusp in ZFC curve at 24 K. These are  characteristic of a spin glass transition, which were also observed in previous studies on the polycrystalline NBMO samples~\cite{Trukhanov:2002p184424,Nakajima:2004p2283}.   Fig~\ref{fig:MT-MH}(b) shows the $M$ vs. $H$ curve of the film  after subtracting the diamagnetic contribution of the STO substrate. A small hysteresis is observed, with a coercive field of 400 Oe. Magnetic hysteresis is observed in ferromagnets, spin glasses, superparamagnets, etc., and Arrott plot  is a conventional way to distinguish ferromagnetism from other phases~\cite{Arrott:1986p3456}. To obtain these we have plotted $M^2$ as a function of $H/M$ in Fig~\ref{fig:MT-MH}(c). The intercept of the extrapolated $M^2$ (denoted by the red line in Fig~\ref{fig:MT-MH}(c)) is on the positive $H/M$ axis, which establishes the absence of any ferromagnetism in our NBMO sample.

The spin glass behavior  below 24 K is further confirmed by  isothermal remanent magnetization (IRM) measurement. For this, the film was cooled to 10 K from room temperature in zero field and a constant magnetic field of 1000 Oe  was applied for 10 minutes. DC moment was recorded as a function of time ($t$) just after switching off the field. The $M_{IRM}$ exhibits a slow decay  [Fig~\ref{fig:MT-MH} (d)], indicative of glassy behaviour. Moreover, $M_{IRM}$ can be fitted using a logarithmic function $M_{IRM} = M_0 - S ln(1+t/t_0)$ ( red curve in Fig~\ref{fig:MT-MH}(d))~\cite{Middey:2011p144419} similar to other spin glasses~\cite{Majumdar:1999p329,Middey:2011p144419}. The viscosity coefficient $S$ is 0.079 emu/cm\textsuperscript{3}.

 In summary, we have grown and investigated epitaxial thin film of disordered NBMO on STO (001) substrate.  A combination of RHEED, AFM, XRR and XRD has confirmed the excellent morphological and structural quality of these films. In sharp contrast to a MIT and ferromagnetism, observed in single crystal of A-site disordered NBMO, our films are insulating and become spin-glass at low temperature. These differences might be related with several factors such as electronic reconstruction due to interfacial polarity mismatch, finite thickness effect etc.~\cite{Garcia:2010p1038,Lee:2010p257204,Bruno:2011p147205} and need to be explored further. This present work also opens the prospect of studying phase competition in Ba-based half-doped manganites through variation of epitaxial strain, interface engineering and geometrical lattice engineering.

 See the supplementary material for the XRD and transport data of a 100 uc thick NBMO film.

The authors acknowledge AFM, XRD and SQUID facilities at the Department of Physics, IISc Bangalore. SM acknowledges DST Nanomission grant  (DST/NM/NS/2018/246) and Infosys Foundation, Bangalore for  financial support.

%

\end{document}